\documentclass[conference]{IEEEtran}
\usepackage[T1]{fontenc}
\usepackage{amsmath}
\usepackage{comment}
\usepackage{amssymb}
\usepackage{amsfonts}
\usepackage{cite}
\usepackage{stfloats}
\usepackage{graphicx}
\usepackage{epsfig}
\usepackage{subfigure}
\usepackage{psfrag}
\usepackage{cite}
\usepackage{latexsym}
\usepackage{url}
\usepackage{color}
\usepackage{multirow}
\usepackage{algorithm}
\usepackage{algorithmic}

\IEEEoverridecommandlockouts
\begin{document}
\title{On the Performance Gain of Integrated Sensing and Communications: A Subspace Correlation Perspective}
\author{Shihang Lu$^{1}$, Xiao Meng$^{1,2}$, Zhen Du$^{3}$, Yifeng Xiong$^{1}$ and Fan Liu$^{1}$ 
\thanks{This work was supported in part by Special Funds for the Cultivation of Guangdong College Students' Scientific and Technological Innovation (``Climbing Program'' Special Funds, No. pdjh2022c0028).} 
\\ 
$^{1}$ Southern University of Science and Technology, Shenzhen, China \\
$^{2}$ Beijing Institute of Technology, Beijing, China\\
$^{3}$ Nanjing University of Information Science and Technology, Nanjing, China \\
E-mail: lush2021@mail.sustech.edu.cn; mengxiao94@bit.edu.cn;  duzhen@nuist.edu.cn; \\ yx8n18@soton.ac.uk; liuf6@sustech.edu.cn
}

\maketitle
\begin{abstract}
In this paper, we shed light on the performance gain of integrated sensing and communications (ISAC) from the perspective of channel correlations between radar sensing and communication (S$\&$C), namely ISAC subspace correlation. To begin with, we consider a multi-input multi-output (MIMO) ISAC system and reveal that the optimal ISAC signal is in the subspace spanned by the transmitted steering vectors of the sensing channel and the right singular matrix of the communication channel. By leveraging this result, we study a basic ISAC scenario with a single target and a single-antenna communication user, and derive the optimal waveform covariance matrix for minimizing the estimation error under a given communication rate constraint. To quantify the integration gain of ISAC systems, we define the subspace ``correlation coefficient'' to characterize the coupling effect between S$\&$C channels. Finally, numerical results are provided to validate the effectiveness of the proposed approaches.
\end{abstract}
\begin{IEEEkeywords}
Integrated sensing and communications, integration gain, signal processing, waveform optimization.
\end{IEEEkeywords}
\section{Introduction}
In recent years, radar sensing and wireless communication systems have been gradually moving from separation to integration \cite{liu2022seventy}. It is foreseeable that integrated sensing and communications (ISAC) would be one of the key enablers of next-generation wireless networks for supporting a variety of emerging applications, such as vehicle-to-everything (V2X) communications and unmanned aerial vehicle (UAV) networks \cite{zhang2021overview,liu_an2022survey, Overview_wuqingqing2021comprehensive}.
The shared utilization of restricted resources, specifically, hardware platform, wireless spectrum as well as energy consumption, bring forward superior benefits for both sensing and communication (S$\&$C). Therefore, ISAC systems can attain significant performance gain over individual S$\&$C systems in terms of the wireless/hardware resource efficiency, 
namely the \textit{integration gain} \cite{cui2021integrating}. 

Intuitively, if S$\&$C systems are completely competing for orthogonal wireless resources (e.g., time slots, spectrum or wireless channels), the integration gain might be very limited. This corresponds to the coexistence scenario where there is only a weak correlation between S$\&$C channels\cite{zhengle2019RCC}. To improve the efficiency of exploiting time, frequency and spatial resources, one may design a unified transmit waveform for achieving S$\&$C functionalities simultaneously \cite{zhonghao2022,lu2022degrees,liu2021dual}. For example, joint ISAC transmit beamforming is able to boost both communication rate and radar estimation accuracy as compared with communication-only and sensing-only design \cite{liu2021cramer,hua2022mimo}. Recent information-theoretic research on ISAC pointed out that the integration gain can be attained via exploiting the potential subspace correlations between S$\&$C \cite{xiong2022flowing}. However, how to quantify the integration gain from the perspective of the subspace correlation still remains open, which motivates our investigation in this paper. 

In this paper, we consider a multi-input and multi-output (MIMO) ISAC system and study the waveform optimization problem that minimizes the Cram\'er-Rao bound (CRB) for target estimation under the communication rate constraint. We first reveal that the optimal ISAC waveform lives in a subspace spanned by both S$\&$C channels. To better understand the subspace correlations, we then focus on a simplistic model with one single target and one single-antenna communication user. As a further step, we optimize the transmit waveform by considering two criteria, namely angle-only (AO) and determinant-maximization (Det-Max) criteria. Moreover, closed-form solutions are derived in both cases to provide insights from the perspective of the subspace correlation. Finally, we define the subspace correlation coefficient to quantify the integration gain in the single-target and single-user scenario, which are validated by numerical simulations.

{\it Notations}: In this paper, lowercase and uppercase boldface letters refer to vectors and matrices, respectively. $\mathbb{C}^{x \times y}$ denotes the spaces of complex matrices with dimension $x \times y$. $\operatorname{span}(\cdot)$ denotes the linear span of a set of vectors. Conjugate, transpose, conjugate transpose, trace, maximum eigenvalue, rank and determinant operators are denoted by $(\cdot)^{\mathsf{*}}$, $(\cdot)^{\mathsf{T}}$, $(\cdot)^{\mathsf{H}}$, $\operatorname{tr}(\cdot)$, $\lambda_{\max}(\cdot)$ and $\operatorname{rank}(\cdot)$ and $\operatorname{det}(\cdot)$, respectively. $\|\cdot\|$ denotes the Euclidean norm. $\mathbb{E}(\cdot)$ denotes the statistical expectation. $\mathcal{CN}(0,1)$ denotes the circularly symmetric complex normal distribution with zero mean. $x \propto y$ denotes that $x$ is proportional to $y$. For a complex number $z$, $|z|$ denotes its modulus and $\angle z$ denotes its phase, respectively.  
\section{System Model}
Let us consider a MIMO ISAC system. The ISAC base station (ISAC-BS) is equipped with $N_t$ transmit antennas and $N_r$ receive antennas, which is assumed to serve an $N_u$-antenna communication user (CU) and detect multiple targets simultaneously. In what follows, we elaborate on the model of S$\&$C.

\subsection{Sensing Model}
We assume that there are $K$ targets to be estimated with unknown angles and reflecting coefficients. The echo signals received at the ISAC-BS can be denoted by 
\begin{align}\label{sensing_signals}
    \mathbf{Y}_{\mathrm{s}} = \sum_{k=1}^{K} \alpha_{k} \mathbf{b}\left(\theta_{k}\right) \mathbf{a}^{\mathsf{T}}\left(\theta_{k}\right) {\mathbf{X}} + \mathbf{Z}_{\mathrm{s}}, 
\end{align}
where $\mathbf{a}(\theta_k)\in \mathbb{C}^{N_{T} \times 1}$ and $\mathbf{b}(\theta_k)  \in \mathbb{C}^{N_{R} \times 1}$  denote the steering vectors for the uniform linear array (ULA), $\mathbf{Z}_{\mathrm{s}} \in  \mathbb{C}^{N_{r} \times T}$ represents the Gaussian noise each column following  $\mathcal{CN}(\mathbf{0},\mathbf{Q})$ in an independent and identically distributed (i.i.d.) manner, $\alpha_k$ denotes the measurable complex reflection coefficient of the $k$-th target, $\theta_k$ is the $k$-th target's azimuth angle of location at a particular range bin of interest, and $\mathbf{X}= [{\mathbf{x}}(1), \ldots ,{\mathbf{x}}(T)]\in \mathbb{C}^{ N_t \times T }$ denotes the dual-functional transmit signal sent by the ISAC-BS, in which $T$ represents the sampling numbers in radar pulse/communication frame, respectively.
For notational convenience, let
$\mathbf{A} = \left[\mathbf{a}\left(\theta_{1}\right), \ldots, \mathbf{a}\left(\theta_{K}\right)\right] $, 
$\mathbf{B}  = \left[\mathbf{b}\left(\theta_{1}\right), \ldots, \mathbf{b}\left(\theta_{K}\right)\right]$, 
and $\boldsymbol{\Sigma} = \mathrm{diag} \left( \alpha_1, \ldots, \alpha_K \right)$, the echo signals can be recast as 
\begin{align}\label{recast_sensing_signals}
    \mathbf{Y}_{\mathrm{s}} = \mathbf{B} \boldsymbol{\Sigma} \mathbf{A}^{\mathsf{T}} \mathbf{X}  + \mathbf{Z}_{\mathrm{s}}.
\end{align}

Let $\boldsymbol{\theta} = \left[\theta_{1}, \ldots, \theta_{K}\right]^\mathsf{T}$ and $\boldsymbol{\alpha} = \left[\alpha_{1}, \ldots, \alpha_{K}\right]^\mathsf{T}$ denote the unknown target parameters to be estimated. Accordingly, the Fisher information matrix (FIM) with respect to $\boldsymbol{\theta}$, and the real and imaginary parts of $\boldsymbol{\alpha}$ in \eqref{sensing_signals} can be expressed in the form of \cite{lijian2007range}
\begin{equation}\label{FIM}
    \mathbf{F}=2\left[\begin{array}{ccc}
        \operatorname{Re}\left(\mathbf{F}_{11}\right) & \operatorname{Re}\left(\mathbf{F}_{12}\right) & -\operatorname{Im}\left(\mathbf{F}_{12}\right) \\
        \operatorname{Re}^{\mathsf{T}}\left(\mathbf{F}_{12}\right) & \operatorname{Re}\left(\mathbf{F}_{22}\right) & -\operatorname{Im}\left(\mathbf{F}_{22}\right) \\
        -\operatorname{Im}^{\mathsf{T}}\left(\mathbf{F}_{12}\right) & -\operatorname{Im}^{\mathsf{T}}\left(\mathbf{F}_{22}\right) & \operatorname{Re}\left(\mathbf{F}_{22}\right)
        \end{array}\right],
\end{equation}
with $\mathbf{F}_{ij}$, $i,j = \{1,2\}$ given by 
\begin{subequations}\label{FIM_elements}
	\begin{align}
		&\mathbf{F}_{11}  = T\left(\dot{\mathbf{B}}^{\mathsf{H}} \mathbf{Q}^{-1} \dot{\mathbf{B}}\right) \odot \left(\boldsymbol{\Sigma}^{*} \mathbf{A}^{\mathsf{H}} \mathbf{R}_{\mathbf{x}}^{\mathsf{T}} \mathbf{A} \boldsymbol{\Sigma} \right) +T\left(\dot{\mathbf{B}}^{\mathsf{H}} \mathbf{Q}^{-1} \mathbf{B}\right) \nonumber\\ &\odot\left(\boldsymbol{\Sigma}^{*} \mathbf{A}^{\mathsf{H}} \mathbf{R}_{\mathbf{x}}^{\mathsf{T}} \dot{\mathbf{A}} \boldsymbol{\Sigma} \right) + T\left({\mathbf{B}}^{\mathsf{H}} \mathbf{Q}^{-1} \dot{\mathbf{B}}\right) \odot \left(\boldsymbol{\Sigma}^{*} \dot{\mathbf{A}}^{\mathsf{H}} \mathbf{R}_{\mathbf{x}}^{\mathsf{T}} \mathbf{A} \boldsymbol{\Sigma} \right) \nonumber \\
		& + T\left({\mathbf{B}}^{\mathsf{H}} \mathbf{Q}^{-1} {\mathbf{B}}\right) \odot \left(\boldsymbol{\Sigma}^{*} \dot{\mathbf{A}}^{\mathsf{H}} \mathbf{R}_{\mathbf{x}}^{\mathsf{T}} \dot{\mathbf{A}} \boldsymbol{\Sigma} \right), \\
		& \mathbf{F}_{12} = T\left(\dot{\mathbf{B}}^{\mathsf{H}} \mathbf{Q}^{-1} {\mathbf{B}}\right) \odot \left(\boldsymbol{\Sigma}^{*} \mathbf{A}^{\mathsf{H}} \mathbf{R}_{\mathbf{x}}^{\mathsf{T}} \mathbf{A} \right) 
		+T\left({\mathbf{B}}^{\mathsf{H}} \mathbf{Q}^{-1} {\mathbf{B}}\right) \nonumber \\
		&\odot \left(\boldsymbol{\Sigma}^{*} \dot{\mathbf{A}}^{\mathsf{H}} \mathbf{R}_{\mathbf{x}}^{\mathsf{T}} \mathbf{A} \right), \\
		& \mathbf{F}_{22} = T\left({\mathbf{B}}^{\mathsf{H}} \mathbf{Q}^{-1} {\mathbf{B}}\right) \odot\left({\mathbf{A}}^{\mathsf{H}} \mathbf{R}_{\mathbf{x}}^{\mathsf{T}} \mathbf{A}\right), 
		\end{align}
\end{subequations}
where $\dot{\mathbf{A}}=[\dot{\mathbf{a}}\left(\theta_{1}\right),  \ldots , \dot{\mathbf{a}}\left(\theta_{K}\right)], 
\dot{\mathbf{B}} =[\dot{\mathbf{b}}\left(\theta_{1}\right),  \ldots , \dot{\mathbf{b}}\left(\theta_{K}\right)]$ and $\mathbf{R}_{\mathbf{x}} =(1/T)\mathbf{X} \mathbf{X}^\mathsf{H}$ denotes the sample covariance matrix.
Therefore, the corresponding Cramér-Rao Bound (CRB) can be expressed as 
\begin{equation}\label{CRB}
    \mathsf{CRB}(\boldsymbol{\theta},\boldsymbol{\alpha})  = \mathrm{tr}(\mathbf{F}^{-1}).
\end{equation}

\subsection{Communication Model}
We consider the point-to-point MIMO communication model in this paper. The received signals of the CU can be denoted by
\begin{equation}
    \mathbf{Y}_{\mathrm{c}} = \mathbf{H}_{\mathrm{c}} {\mathbf{X}} + \mathbf{Z}_{\mathrm{c}},
\end{equation}
where $\mathbf{H}_{\mathrm{c}} \in \mathbb{C}^{ N_u \times N_t }$ denotes the MIMO communication channel matrix, and $\mathbf{Z}_{\mathrm{c}}$ denotes the noise with zero mean and variance $\sigma_c^2$.
Therefore, the achievable communication rate can be expressed as
\begin{align}
    \mathsf{C}  = \log \det \left(  \mathbf{I} + \sigma_c^{-2} \mathbf{H}_{\mathrm{c}} \widetilde{\mathbf{R}}_{\mathrm{x}} \mathbf{H}_{\mathrm{c}}^{\mathsf{H}} \right),
\end{align}
where $\mathbf{I}$ denotes the identity matrix with dimension $N_u$ and $\widetilde{\mathbf{R}}_{\mathrm{x}} \approx (1/T) \mathbf{X} \mathbf{X}^H$ becomes accurate when $T$ is sufficiently large. 

\subsection{Problem Formulation and Optimal Waveform Covariance Matrix}
To achieve a favorable performance tradeoff between S$\&$C, $\mathbf{R}_{\mathrm{x}}$ should be conceived to minimize the CRB with respect to the requirement of the communication rate. In the multi-target scenario with fixed $\boldsymbol{\theta}$ and $\boldsymbol{\alpha}$, the waveform optimization problem is formulated as 
\begin{subequations}  \label{P1}
	\begin{align}
        \min_{ \mathbf{R}_{\mathrm{x}} } &~~ \mathsf{CRB}(\boldsymbol{\theta},\boldsymbol{\alpha}) \nonumber\\
        \label{Comm_Cons}
        ~~~\mathrm{s.t.} &~~ \mathsf{C} \geq \gamma, \\
        &~~ \mathbf{R}_{\mathrm{x}} \succeq \mathbf{0},~\mathrm{tr}\left(\mathbf{R}_{\mathrm{x}}\right) = P,
\end{align}
\end{subequations}
where $\gamma$ is a required rate threshold of the CU, and $P$ is the total transmit power budget. To show the optimal structure of the transmitted waveform in problem \eqref{P1}, we have the following theorem.

{\it \textbf{Theorem 1 (Structure of the Optimal Waveform):}}
The optimal $\mathbf{R}_{\mathrm{x}}$ can be written in the form as
    \begin{equation}
        {\mathbf{{R}}}_{\mathbf{x}}^{\star} = \mathbf{U} \boldsymbol{\Lambda} {\mathbf{U}}^\mathsf{H},
    \end{equation}
where $ \mathbf{U} = [ \mathbf{A}^{*}~\dot{\mathbf{A}}^{*}~\mathbf{V}_{\mathrm{c}} ] \in \mathbb{C}^{ N_t \times (2K + \mathrm{rank}(\mathbf{H}_{\mathrm{c}}))}$ and $\boldsymbol{\Lambda}$ is a positive semi-definite matrix. The matrix $\mathbf{V}_{\mathrm{c}}$ is the right singular vector matrix based on the singular value decomposition (SVD) $\mathbf{H}_{\mathrm{c}} = \mathbf{U}_{\mathrm{c}} \boldsymbol{\Sigma}_{\mathrm{c}} \mathbf{V}_{\mathrm{c}}^\mathsf{H}$. 

{\it Proof:} Let $\mathbf{R}_{\mathrm{x}} = \boldsymbol{\Delta}\boldsymbol{\Delta}^\mathsf{H}$. We can decompose $\boldsymbol{\Delta}$ as $\boldsymbol{\Delta} = \mathbf{P}_U \boldsymbol{\Delta} + \mathbf{P}_U^{\bot} \boldsymbol{\Delta}$, where $\mathbf{P}_U$ represent the orthogonal projector onto the subspace spanned by the columns of $\mathbf{U}$ and $\mathbf{P}_U^{\bot} = \mathbf{I}-\mathbf{P}_U$. It can be similarly confirmed that the orthogonal part $\mathbf{P}_U^{\bot} \boldsymbol{\Delta} = \mathbf{0}$ still holds at the optimum for CRB-only optimization problem even with the rate constraint \cite{lijian2007range}. Due to the strict page limit, we refer readers to [13, App. C] for more details. 

The remaining task is to confirm that the orthogonal part does not affect the achievable rate $\mathsf{C}$. This is true since $\mathbf{P}_U^{\bot}\mathbf{V}_{\mathrm{c}} = \mathbf{0}$, which indicates that the orthogonal part has no contribution to the achievable rate. Hence, we have
\begin{equation}
    {\mathbf{{R}}}_{\mathbf{x}}^{\star} = \mathbf{P}_U \boldsymbol{\Delta} \boldsymbol{\Delta}^\mathsf{H} \mathbf{P}_U \triangleq \mathbf{U} \boldsymbol{\Lambda} {\mathbf{U}}^\mathsf{H},
\end{equation}
which completes the proof.
$\hfill\blacksquare$

We remark that {\it \textbf{Theorem 1}} indicates the optimal ${\mathbf{{R}}}_{\mathbf{x}}^{\star}$ belongs to the subspace spanned by the columns of $[ \mathbf{A}^{*}~\dot{\mathbf{A}}^{*} ]$ and $\mathbf{V}_{\mathrm{c}}$. To avoid confusion, we define $\operatorname{span}\{\mathbf{A}^{*}, \dot{\mathbf{A}}^{*}\}$ and $\operatorname{span}\{ \mathbf{V}_{\mathrm{c}} \}$ as the {\it sensing subspace} and {\it communication subspace}, respectively, and define the subspace spanned by the columns of $\mathbf{U}$ as the {\it ISAC subspace}. Intuitively, if there is a strong correlation between S$\&$C subspaces, the ISAC systems might attain better integration gain. In the following section, we will study a simplistic model to illustrate this point.     

\section{Case Study}\label{Sec3}
With the above observations in mind, let us consider the scenario with one target and one single-antenna communication user. The echo signals received at the ISAC-BS can be simplified by
\begin{align}
    \mathbf{Y}_{\mathrm{s}} = \alpha \mathbf{b}\left(\theta\right) \mathbf{a}^{\mathsf{T}}\left(\theta\right) {\mathbf{X}} + \mathbf{Z}_{\mathrm{s}},
\end{align}
In order to gain further insights, we assume that
1) the noise is white Gaussian distributed, i.e., $\mathbf{Q} = \sigma_s^2 \mathbf{I}$, where $\sigma_s^2$ denotes the average interference power;
2) the reference point for the sensing transmit as well as the receive linear array is chosen so that $\mathbf{a}^\mathsf{H} \dot{\mathbf{a}} = 0$ and $\mathbf{b}^\mathsf{H} \dot{\mathbf{b}} = 0$.
One may therefore express each entry of the FIM as
\begin{subequations}\label{FIM_Elements}
    \begin{align}
        &\mathsf{F}_{11} 
         = T |\alpha|^2 \sigma_s^{-2}\left(  \| \dot{\mathbf{b}} \|^2 \operatorname{tr}( \mathbf{a}^* \mathbf{a}^\mathsf{T} \mathbf{R}_{\mathbf{x}}) + \operatorname{tr}( \dot{\mathbf{a}} ^ * \dot{\mathbf{a}}^\mathsf{T} \mathbf{R}_{\mathbf{x}}) \right), \\
         & \mathsf{F}_{12} 
        = T \alpha^{*}  \sigma_s^{-2} \operatorname{tr} ( \dot{\mathbf{a}}^*  {\mathbf{a}}^\mathsf{T} \mathbf{R}_{\mathbf{x}}), \\ 
        & \mathsf{F}_{22} 
         = T \sigma_s^{-2} \operatorname{tr}(\mathbf{a}^* \mathbf{a}^\mathsf{T} \mathbf{R}_{\mathbf{x}}).
    \end{align}
\end{subequations}

By denoting the communication channel vector as $\mathbf{h}_{\mathrm{c}}\in \mathbb{C}^{N_t \times 1}$, the communication signal $\mathbf{y}_{\mathrm{c}}^{\mathsf{T}} \in \mathbb{C}^{1 \times T}$ received at the CU is  
\begin{equation}
        \mathbf{y}_{\mathrm{c}}^{\mathsf{T}} = \mathbf{h}_{\mathrm{c}}^\mathsf{H} {\mathbf{X}} + \mathbf{z}_{\mathrm{c}}^{\mathsf{T}}.
\end{equation}
Accordingly, the communication rate constraint in \eqref{Comm_Cons} can be recast as
\begin{equation}
    \mathrm{tr}(\mathbf{Q}_{\mathrm{c}} \mathbf{R}_{\mathrm{x}}) \ge \Gamma,
\end{equation} 
where $\Gamma = (2^{\gamma}-1)\sigma_c^2$ denotes the required  signal-to-noise ratio (SNR) and $\mathbf{Q}_{\mathrm{c}} = \mathbf{h}_{\mathrm{c}} \mathbf{h}_{\mathrm{c}}^\mathsf{H}$. By straightforward application of the Schur complement condition \cite{zhang2006schur}, the non-convex optimization problem \eqref{P1} may be recast into a semidefinite program (SDP) as
\begin{subequations}\label{P2}
	\begin{align}
		\min_{ \{t_k\}_{k=1}^{3}, \mathbf{R}_{\mathrm{x}} } &~~ \sum_{k=1}^{3}t_k  \nonumber\\
        \label{SDP_Cons}
		~~~\mathrm{s.t.} ~~ & \left[\begin{array}{cc}
			\mathbf{F} & \mathbf{e}_k \\
			\mathbf{e}_k^{\mathsf{T}} & t_k
			\end{array}\right] \succeq \mathbf{0}, \quad k=1, 2, 3,  \\ 
			&~~ \mathrm{tr}(\mathbf{Q}_{\mathrm{c}} \mathbf{R}_{\mathrm{x}}) \ge \Gamma, \\
		&~~ \mathbf{R}_{\mathrm{x}} \succeq \mathbf{0},~\mathrm{tr}\left(\mathbf{R}_{\mathrm{x}}\right) = P, 
	\end{align}
\end{subequations} 
where $\mathbf{F}$ is the FIM given in \eqref{FIM} with each entry in \eqref{FIM_Elements}, $\{t_k\}_{k=1}^{3}$ are auxiliary variables and $\mathbf{e}_k$ is the $k$-th column of the identity matrix of dimension three, respectively.


Notice that problem \eqref{P2} can be solved via numerical tools like CVX \cite{boyd2004convex}. To futher simplify the problem, we will show that closed-form solutions are attainable with the aid of the following lemma.

{\it \textbf{Lemma 1:}}
The optimal $\mathbf{R}_{\mathrm{x}}$ of problem \eqref{P2} is rank-1 and thus it can be expressed as
\begin{align}\label{Lemma_Rx_opt}
	\mathbf{R}_{\mathrm{x}} = \mathbf{u}\mathbf{u}^\mathsf{H}, \mathbf{u} = \nu_1 \mathbf{a}_{\mathrm{u}} + \nu_2  \mathbf{a}_{\mathrm{d}} +  \nu_3 \mathbf{a}_{\mathrm{h}},
\end{align} 
where 
\begin{align}\label{orthogonal_basis}
    \mathbf{a}_{\mathrm{u}} = \mathbf{a}^*,
    \mathbf{a}_{\mathrm{d}} = \frac{\dot{\mathbf{a}}^* }{\|\dot{\mathbf{a}}^* \|}, \mathbf{a}_{\mathrm{h}} = \frac{  \mathbf{h}_{\mathrm{c}} - \mathbf{a}_{\mathrm{u}}^{\mathsf{H}} \mathbf{h}_{\mathrm{c}} \mathbf{a}_{\mathrm{u}} - \mathbf{a}_{\mathrm{d}}^{\mathsf{H}} \mathbf{h}_{\mathrm{c}} \mathbf{a}_{\mathrm{d}}}{\| \mathbf{h}_{\mathrm{c}} - \mathbf{a}_{\mathrm{u}}^{\mathsf{H}} \mathbf{h}_{\mathrm{c}} \mathbf{a}_{\mathrm{u}} - \mathbf{a}_{\mathrm{d}}^{\mathsf{H}} \mathbf{h}_{\mathrm{c}} \mathbf{a}_{\mathrm{d}} \|},
\end{align}
and $\{\nu_{\ell}\}_{{\ell}=1}^3$ are weighted parameters to be optimized.

{\it Proof:} 
The rank-1 property of the optimal $\mathbf{R}_{\mathrm{x}}$ can be proved by analyzing the Karush-Kuhn-Tucker (KKT) conditions of the problem. Due to the strict page limit, we do not show the detailed procedure and refer readers to \cite{liu2021cramer} for more details. Following the {\it \textbf{Threorem 1}}, it can be readily confirmed that $\mathbf{u} \in \mathrm{span}\{ \mathbf{a}^*, \dot{\mathbf{a}}^*, \mathbf{h}_{\mathrm{c}} \}$. Then we can denote $\mathbf{u}$ by the orthogonal basis $\{\mathbf{a}_{\mathrm{u}}, \mathbf{a}_{\mathrm{d}}, \mathbf{a}_{\mathrm{d}}\}$, which is expressed as \eqref{Lemma_Rx_opt}. This completes the proof.  
$\hfill\blacksquare$

To better harness the main results in {\it \textbf{Lemma 1}}, we present the following two criteria, namely the AO criterion and Det-Max criterion.

\subsection{Angle-Only Criterion}
The angle-only CRB refers to the first entry in the diagonal of the inverse FIM. By leveraging the Schur complement into the FIM, we have
\begin{align}
    \mathsf{CRB}({\theta}) = \mathrm{0.5} \mathsf{F}_{11.2}^{-1},
\end{align}
where $\mathsf{F}_{11.2}^{-1} \triangleq \mathsf{F}_{11}-\mathsf{F}_{12} \mathsf{F}_{22}^{-1} \mathsf{F}_{12}^*$ with each term given in \eqref{FIM_Elements}. By taking $\mathbf{R}_{\mathrm{x}} = \mathbf{u}\mathbf{u}^\mathsf{H}$ of \eqref{Lemma_Rx_opt} into the above equation, we can derive the angle-only CRB as 
\begin{align}\label{CRB_19}
    \mathsf{CRB}({\theta}) = \frac{\sigma_s^{2} }{ 2 T |\alpha|^2  |\nu_1|^2 \| \dot{\mathbf{b}} \|^2 }.
\end{align}
Accordingly, the CRB optimization problem in angle-only criterion can be recast as
\begin{subequations}\label{P_AngleOnly}
	\begin{align}
		\max_{ \{\nu_{\ell}\}_{{\ell}=1}^{3} } &~~  2 T |\alpha|^2  |\nu_1|^2 \| \dot{\mathbf{b}} \|^2/\sigma_s^2 \nonumber\\\label{P_AngleOnly_C1}
		~~~\mathrm{s.t.} ~~ & ~~ |\nu_1 \mathbf{h}_{\mathrm{c}}^\mathsf{H} \mathbf{a}_{\mathrm{u}} + \nu_2 \mathbf{h}_{\mathrm{c}}^\mathsf{H} \mathbf{a}_{\mathrm{d}} + \nu_3 \mathbf{h}_{\mathrm{c}}^\mathsf{H} \mathbf{a}_{\mathrm{h}}|^2 \ge \Gamma, \\\label{P_AngleOnly_C2} 
		&~~ |\nu_1|^2 + |\nu_2|^2 + |\nu_3|^2 = P.
\end{align}
\end{subequations}
It can be readily confirmed that \eqref{P_AngleOnly_C1} holds active at the optimum. Notice that by aligning the phase of each $\{\nu_{\ell}^{\star}\}_{{\ell}=1}^{3}$ with the corresponding coefficient in the left-hand side of \eqref{P_AngleOnly_C1}, the objective value keeps unchanged without violating any constraints.  Therefore, the phases of each optimal $\{\nu_{\ell}^{\star}\}_{{\ell}=1}^{3}$ can be given by
\begin{align}
    \zeta_1 = -\angle \mathbf{h}_{\mathrm{c}}^\mathsf{H} \mathbf{a}_{\mathrm{u}},~\zeta_2 = -\angle \mathbf{h}_{\mathrm{c}}^\mathsf{H} \mathbf{a}_{\mathrm{d}},~\zeta_3 = -\angle \mathbf{h}_{\mathrm{c}}^\mathsf{H} \mathbf{a}_{\mathrm{h}},
\end{align}
which implies $\nu_{\ell}^{\star} = |\nu_{\ell}^{\star}| e^{j\zeta_{\ell}}, {\ell} =1,2,3$. The optimization problem \eqref{P_AngleOnly} may be now equivalently formulated as
\begin{subequations}\label{P_AngleOnly_Abs}
	\begin{align}
		\max_{ \{|\nu_{\ell}|\}_{{\ell}=1}^{3} } &~~  2 T |\alpha|^2  |\nu_1|^2 \| \dot{\mathbf{b}} \|^2 /\sigma_s^2  \nonumber\\
		~~~\mathrm{s.t.} & ~~ |\nu_1 \mathbf{h}_{\mathrm{c}}^\mathsf{H} \mathbf{a}_{\mathrm{u}} + \nu_2 \mathbf{h}_{\mathrm{c}}^\mathsf{H} \mathbf{a}_{\mathrm{d}} + \nu_3 \mathbf{h}_{\mathrm{c}}^\mathsf{H} \mathbf{a}_{\mathrm{h}}|^2 = \Gamma, \\ 
		&~~ |\nu_1|^2 + |\nu_2|^2 + |\nu_3|^2 = P.
	\end{align}
\end{subequations}
By applying the Lagrange multiplier, the optimal $\{|\nu_k|\}_{k=1}^{3}$ of optimization problem \eqref{P_AngleOnly_Abs} can be derived as \eqref{optimal_v}, shown at the top of this page.
\begin{figure*}[ht]
    \begin{equation}
        \begin{aligned}
            &|\nu_1^{\star}| = \frac{\sqrt{\Gamma} - \varsigma |\nu_3^{\star}| }{ |\mathbf{h}_{\mathrm{c}}^\mathsf{H} \mathbf{a}_{\mathrm{u}}| } = \frac{(1-\varepsilon) \sqrt{\Gamma}+\varsigma \sqrt{-\varpi \Gamma+\frac{P}{A}}}{ |\mathbf{h}_{\mathrm{c}}^\mathsf{H} \mathbf{a}_{\mathrm{u}}|}, 
            ~~|\nu_2^{\star}|  = \frac{|\mathbf{h}_{\mathrm{c}}^\mathsf{H} \mathbf{a}_{\mathrm{d}}| }{| \mathbf{h}_{\mathrm{c}}^\mathsf{H} \mathbf{a}_{\mathsf{h}}|}|\nu_3^{\star}|, 
            ~~|\nu_3^{\star}| = \frac{-B-\sqrt{B^2 - 4AC}}{2A}, \\
            & \varsigma  = \frac{|\mathbf{h}_{\mathrm{c}}^\mathsf{H} \mathbf{a}_{\mathrm{d}}|^2 + | \mathbf{h}_{\mathrm{c}}^\mathsf{H} \mathbf{a}_{\mathrm{h}}|^2}{| \mathbf{h}_{\mathrm{c}}^\mathsf{H} \mathbf{a}_{\mathrm{d}}|}, \varepsilon = \frac{\varsigma^2}{A\left|\mathbf{h}_{\mathrm{c}}^{\mathsf{H}} \mathbf{a}_{\mathrm{u}}\right|^2},
            \varpi = \frac{(1-\varepsilon) }{A\left|\mathbf{h}_{\mathrm{c}}^{\mathrm{H}} \mathbf{a}_{\mathrm{u}}\right|^2}
            A = \frac{\varsigma^2}{|\mathbf{h}_{\mathrm{c}}^\mathsf{H} \mathbf{a}_{\mathrm{u}}|^2}  
            + \frac{|\mathbf{h}_{\mathrm{c}}^\mathsf{H} \mathbf{a}_{\mathrm{d}}|^2}{| \mathbf{h}_{\mathrm{c}}^\mathsf{H} \mathbf{a}_{\mathrm{h}}|^2} + 1,
            B = -\frac{2\varsigma \sqrt{\Gamma}}{|\mathbf{h}_{\mathrm{c}}^\mathsf{H} \mathbf{a}_{\mathrm{u}}|^2}, 
            C = \frac{\Gamma}{|\mathbf{h}_{\mathrm{c}}^\mathsf{H} \mathbf{a}_{\mathrm{u}}|^2} - P.
    \end{aligned}
    \label{optimal_v}
    \end{equation}
    \hrulefill
\end{figure*}

Based on the above derivations, we can reconstruct the optimal $\mathbf{R}_{\mathrm{x}}^{\star}$ in a closed form, which is expressed as 
\begin{align}\label{opt_Rx_AngleOnly}
        \mathbf{R}_{\mathrm{x}}^{\star} = \hat{\mathbf{u}}\hat{\mathbf{u}}^{ \mathsf{H}}, \hat{\mathbf{u}}= |\nu_1^{\star}|e^{j\zeta_1}\mathbf{a}_{\mathrm{u}} + |\nu_2^{\star}|e^{j\zeta_2}\mathbf{a}_{\mathrm{d}} + |\nu_3^{\star}|e^{j\zeta_3}\mathbf{a}_{\mathrm{h}}.
\end{align}

\subsection{Determinant-Maximization Criterion}  
The CRB matrix represents the covariance matrix of those parameters $\{\theta, \operatorname{Re}(\alpha), \operatorname{Im}(\alpha)\}$, which is the inverse of the FIM. Moreover, the determinant of a covariance matrix is also known as a generalized variance. Therefore, minimizing the determinant of the CRB matrix is equivalent to maximizing the determinant of the FIM matrix $|\mathbf{F}|$, which accordingly improves the target estimation performance \cite{lijian2007range,hua2022mimo}. 

In the single-target scenario, we formulate the determinant-maximization optimization problem as 
\begin{subequations}\label{P1_Det}
	\begin{align}
		\max_{ \mathbf{R}_{\mathrm{x}} } &~~  |\mathbf{F}| \nonumber\\
		~~~\mathrm{s.t.} &~~ \mathrm{tr}(\mathbf{Q}_{\mathrm{c}} \mathbf{R}_{\mathrm{x}}) \ge \Gamma, \\
        &~~ \mathbf{R}_{\mathrm{x}} \succeq \mathbf{0},~\mathrm{tr}\left(\mathbf{R}_{\mathrm{x}}\right) = P.
	\end{align}
\end{subequations} 
We remark that problem \eqref{P1_Det} is convex and it can be directly solved by numerical tools \cite{boyd2004convex}. To gain further insights, we leverage the main results in {\it \textbf{Lemma 1}} to attain the optimal solution in the following.

By substituting $\mathbf{R}_{\mathrm{x}} = \mathbf{u}\mathbf{u}^\mathsf{H}$ of \eqref{Lemma_Rx_opt} into the determinant $|\mathbf{F}|$
\begin{align}
    |\mathbf{F}| & = 8 \mathsf{F}_{11.2}\left|\left[\begin{array}{cc}\operatorname{Re}\left(\mathsf{F}_{22}\right) & -\operatorname{Im}\left(\mathsf{F}_{22}\right) \\ -\operatorname{Im}\left(\mathsf{F}_{22}\right) & \operatorname{Re}\left(\mathsf{F}_{22}\right)\end{array}\right]\right| \nonumber \\
    & = 8 T^3 |\alpha|^2 \sigma_s^{-6}   |\nu_1|^6 \| \dot{\mathbf{b}} \|^2  /\sigma_s^2,
\end{align}
we can simplify the determinant-maximization optimization problem as 
\begin{subequations}\label{P_Det}
    \begin{align}
		\max_{ \{|\nu_{\ell}|\}_{\ell=1}^{3} } &~~  8 T^3 |\alpha|^2 \sigma_s^{-6}   |\nu_1|^6 \| \dot{\mathbf{b}} \|^2  /\sigma_s^2 \nonumber \\
		~~~\mathrm{s.t.} & ~~  |\nu_1 \mathbf{h}_{\mathrm{c}}^\mathsf{H} \mathbf{a}_{\mathrm{u}} + \nu_2 \mathbf{h}_{\mathrm{c}}^\mathsf{H} \mathbf{a}_{\mathrm{d}} + \nu_3 \mathbf{h}_{\mathrm{c}}^\mathsf{H} \mathbf{a}_{\mathrm{h}}|^2 = \Gamma, \\ 
		&~~ |\nu_1|^2 + |\nu_2|^2 + |\nu_3|^2 = P.
\end{align}
\end{subequations}
It can be readily observed that the optimization problem \eqref{P_Det} shows the same structure as that of the optimization problem \eqref{P_AngleOnly_Abs}, which indicates that the optimal $\mathbf{R}_{\mathrm{x}}^{\star}$ can be expressed as the same as \eqref{opt_Rx_AngleOnly}.

\section{Discussions About the ISAC Subspace}
\begin{figure}[htbp]
    \centering
    \epsfxsize=1\linewidth
    \includegraphics[width=\columnwidth]{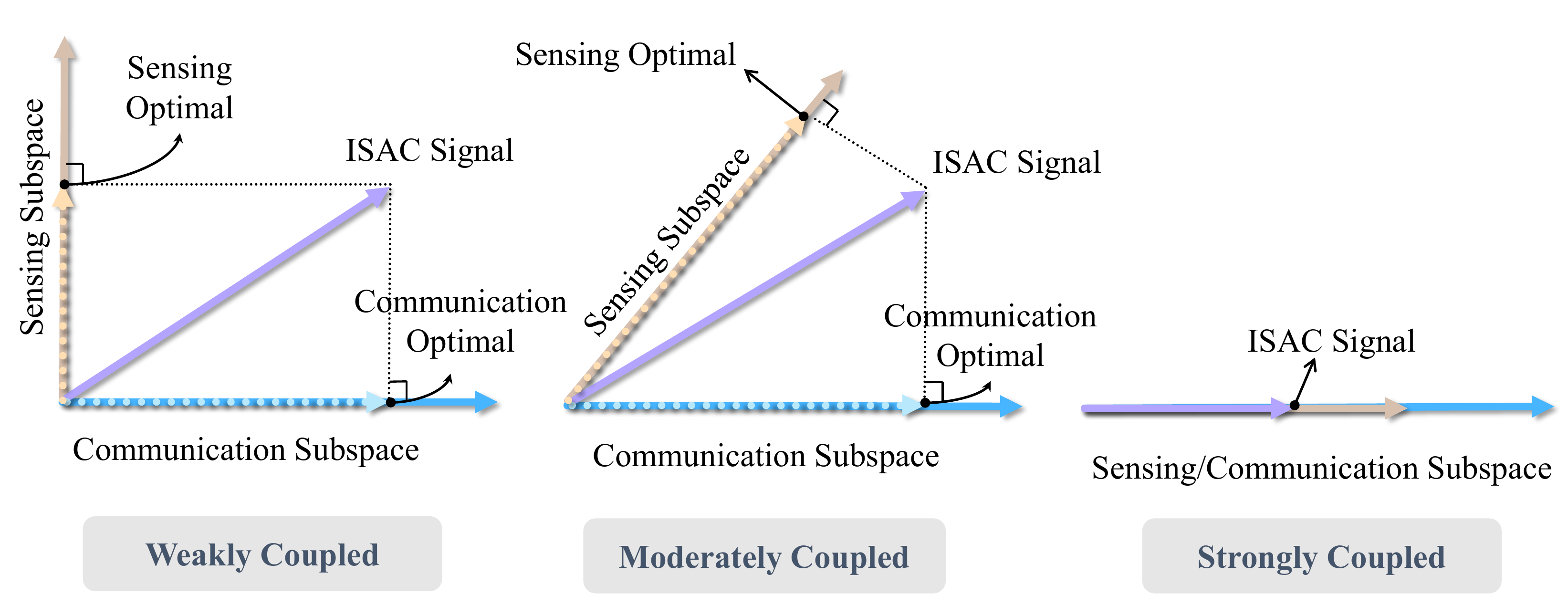}
    \caption{The illustration of the ISAC subspace and its coupling effect.}
    \label{Subspace}
\end{figure}

In ISAC systems, the wireless resources are shared between S$\&$C functionalities, leading to an inherent tradeoff between S$\&$C subspaces \cite{liu2022seventy,xiong2022flowing}. To be specific, if the S$\&$C subspaces are strongly correlated, more signaling resources may be reused between both functionalities, resulting in higher resource efficiency and integration gain in ISAC systems. However, it is difficult to fully characterize the integration gain in a mathematically tractable manner. In this section, we first illustrate the performance gain of ISAC systems originates form the correlation between S$\&$C subspaces intuitively in Fig. \ref{Subspace}. As a step further, we define a ``correlation coefficient'' to qualify the integration gain.

Following the main results in {\it \textbf{Theorem 1}}, the ISAC signal should belong to the ISAC subspace that is spanned by the sensing subspace and communication subspace. The correlation between two subspaces can be characterized by their intersection angle as illustrated in Fig. \ref{Subspace}. The projection of the ISAC signal onto the corresponding subspaces represents the contribution on S$\&$C performance respectively. In the weakly coupled case, Fig. \ref{Subspace} shows that the projection strength reduces to the minimum since S$\&$C subspaces are nearly orthogonal to each other. In the moderately coupled case, the projection onto sensing subspace becomes larger, leading to a more favorable S$\&$C tradeoff compared to its weakly coupled counterpart. Finally, in the strongly coupled case, since the two subspaces are fully aligned, the ISAC signal power can be fully reused to simultaneously boost the performance of S$\&$C. 

\begin{figure}[htbp]
    \centering
    \epsfxsize=1\linewidth
    \includegraphics[width=8.5cm]{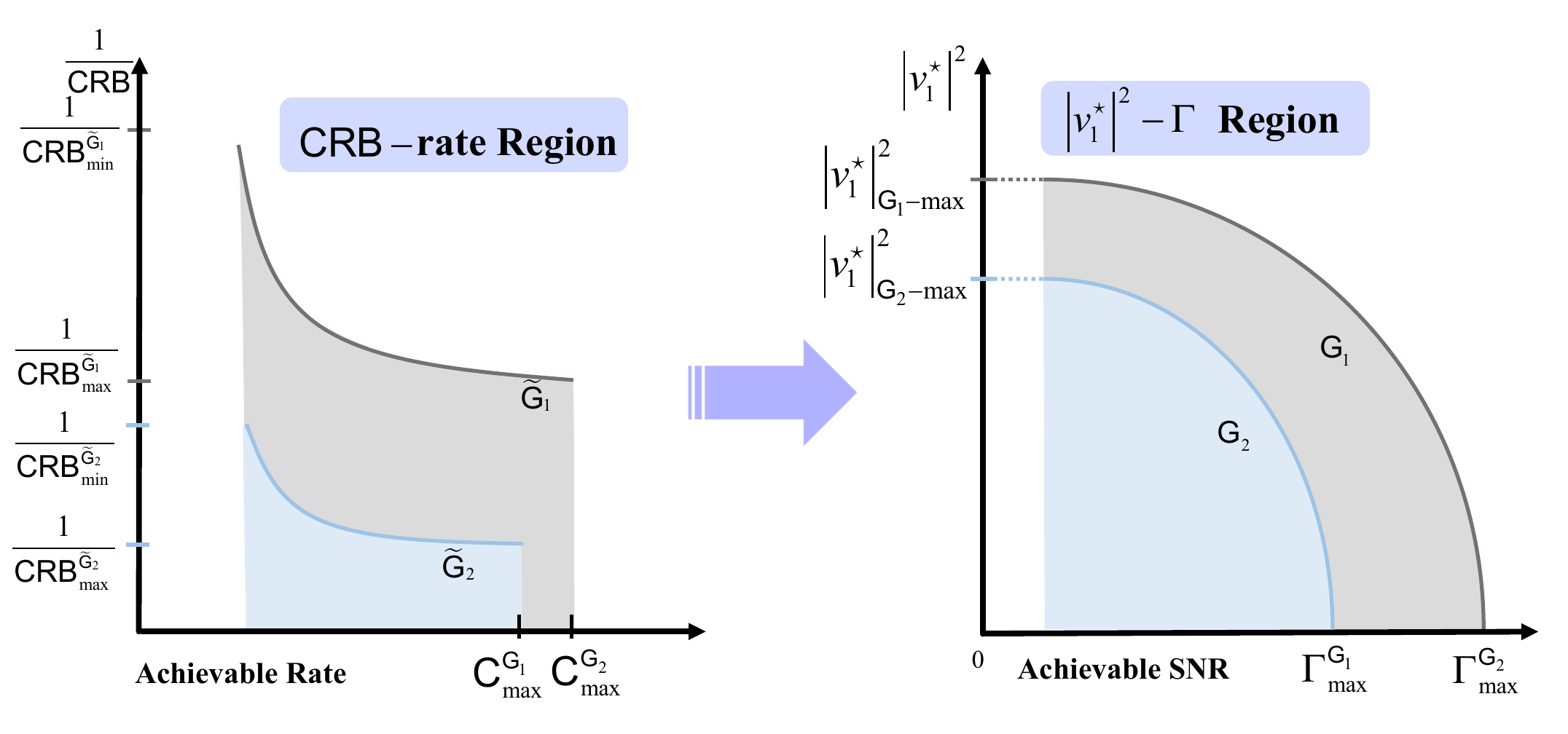}
    \caption{The illustration of the CRB-C region into ``$\left|\nu_1^{\star}\right|^2 - \Gamma$'' region.}
    \label{Reshape}
\end{figure}

The ISAC performance may be characterized as the achievable CRB-rate region \cite{hua2022mimo,liu2021cramer}. A potential metric to evaluate the integration gain is to measure the area enclosed by the Pareto boundary of the CRB-rate region, which accounts that an ISAC system attains high integration gain given a large area. However, it is very difficult to obtain its analytical solutions. To this end, we propose an alternative region area to characterize the ISAC performance gain analytically. It can be observed that given limited resources, the larger communication SNR/rate requirement might lead to larger CRB \cite{hua2022mimo,liu2021cramer}, i.e., $\Gamma  \propto \gamma \propto \mathsf{CRB}$. We further note that the weighted coefficient $\nu_1$ contributes to both S$\&$C in problems \eqref{P_AngleOnly} and \eqref{P_Det}. Given a fixed SNR requirement $\Gamma$, if more transmit power budget is allocated to $\nu_1$, the ISAC system will have better estimation performance since $\mathsf{CRB} \propto 1/ |\nu_1^{\star}|^2$. Based on the above, the following proportional relationship holds true
\begin{align}
    \Gamma \propto \mathsf{CRB} \propto \frac{1}{\left|\nu_1^{\star}\right|^2},
\end{align}
where $|\nu_1^{\star}| ^2$ is given in \eqref{v1_Square}. This states that if the area enclosed by the ``$\left|\nu_1^{\star}\right|^2 - \Gamma$'' Pareto boundary is also large, the corrsponding CRB-rate region area is large. Accordingly, CRB-rate region can be equivalently represented by the ``$\left|\nu_1^{\star}\right|^2 - \Gamma$'' region as illustrated in Fig. \ref{Reshape}. Intuitively, one may characterize the correlation between S$\&$C subspaces by the area of the  ``$\left|\nu_1^{\star}\right|^2 - \Gamma$'' region, defined as the ``correlation coefficient''
\begin{align}\label{Def_G}
    \mathsf{G} = \int_{\Gamma_{1}}^{\Gamma_{2}} \left|\nu_1^{\star}\right|^2 \,d \Gamma, 
\end{align}
which can be computed analytically by \eqref{integration} at the top of the next page, with $0 \leq \Gamma_1 \leq \Gamma_2 \leq \Gamma_{\max} $ and $\Gamma_{\max}$ denoting the maximum achievable SNR.

In a nutshell, an ISAC system with single target and single CU attains high integration gain given a large ``correlation coefficient'' $\mathsf{G}$. In the simulations, we will use the normalized $\mathsf{G}$ calculated by $\mathsf{G}_i = \mathsf{G}_i / \max(\mathsf{G}_i)$ to quantify the integration gain over several ISAC systems, where $i = 1,2,\ldots,i_{m}$ and $i_{m}$ is the channel realization number of times used to denote different ISAC systems. 

\begin{figure}[!h]
    \centering
    \epsfxsize=1\linewidth
    \includegraphics[width=8cm]{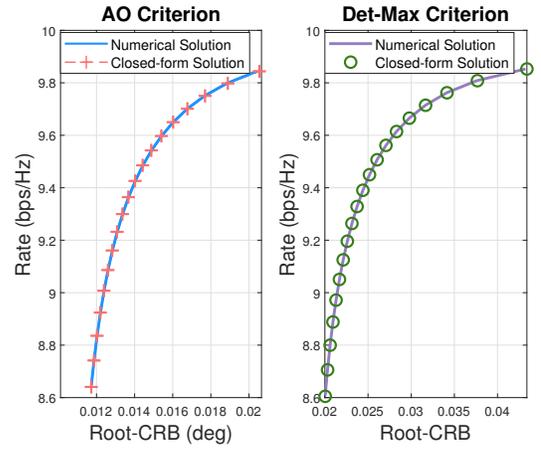}
    \caption{Closed-form and numerical solutions for performance region.}
    \label{Fig1a}
\end{figure}  

\begin{figure}[!t]
    \centering
    \epsfxsize=1\linewidth
    \includegraphics[width=8cm]{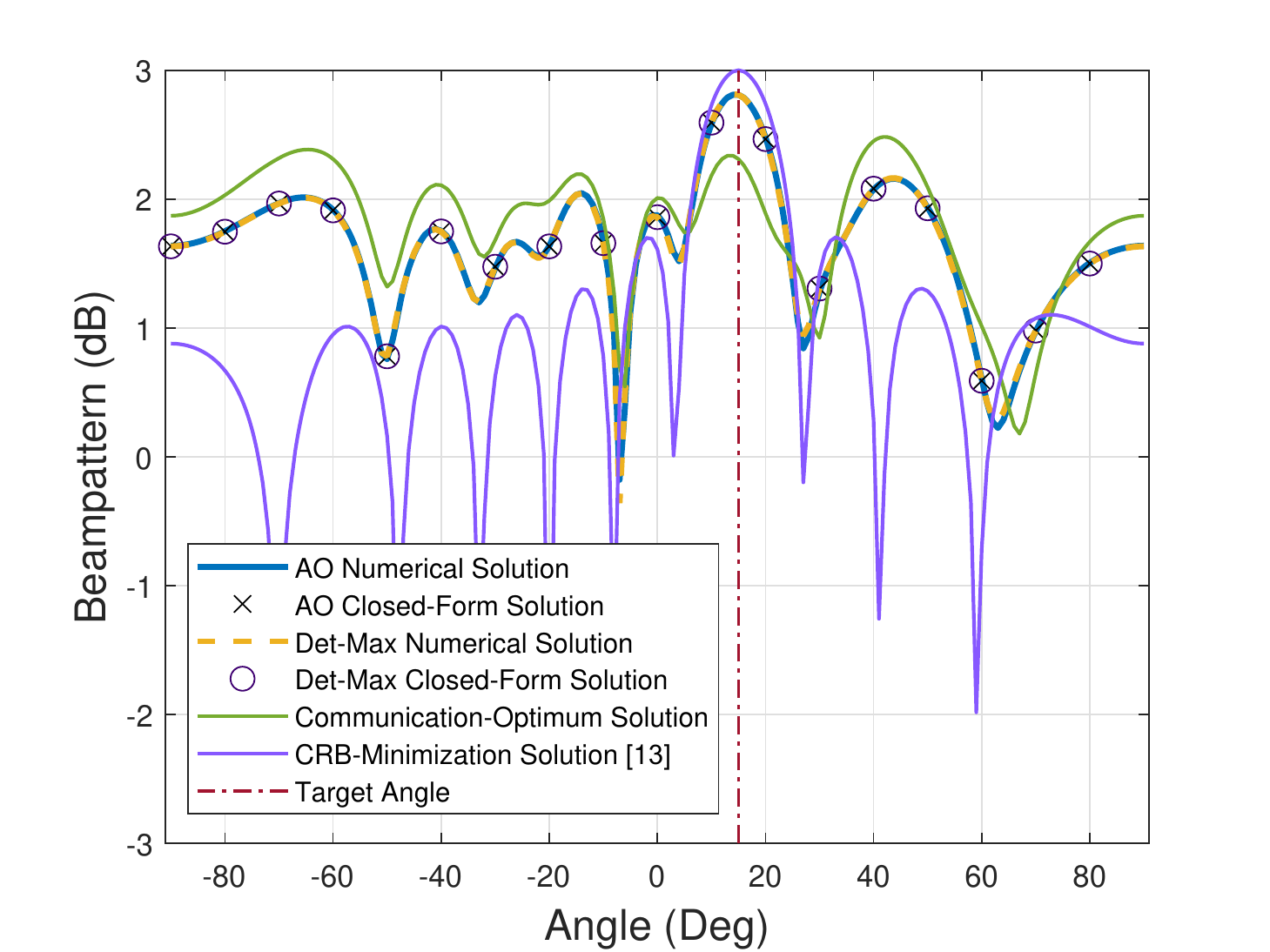}
    \caption{Beampatterns.}
    \label{Fig1b}
\end{figure} 

\begin{figure*}[!t]
    \begin{small}
    \begin{subequations}
        \begin{align}
            \label{v1_Square}
            \left|\nu_1^{\star}\right|^2 &= \left[(1-\varepsilon)^2 \Gamma+\varsigma^2\left(\frac{(\varepsilon-1) \Gamma}{A\left|\mathbf{h}_{\mathrm{c}}^{\mathsf{H}} \mathbf{a}_{\mathrm{u}}\right|^2}+\frac{P}{A}\right)+2 \varsigma(1-\varepsilon) \sqrt{\frac{(\varepsilon-1) \Gamma^2}{A\left|\mathbf{h}_{\mathrm{c}}^H \mathbf{a}_{\mathrm{u}}\right|^2}+\frac{P \Gamma}{A}}\right]/ | \mathbf{h}_{\mathrm{c}}^\mathsf{H} \mathbf{a}_{\mathrm{u}}|^2, \\
            \label{integration}
            \mathsf{G}&= \left[\frac{1}{2}\left((1-\varepsilon)^2+\varpi \varsigma^2\right) \Gamma^2+\frac{P \varsigma^2}{A} \Gamma+2 \varsigma(1-\varepsilon)\left(\frac{2 \varpi \Gamma-\frac{P}{A}}{4 \varpi} \sqrt{\varpi \Gamma^2+\frac{P \Gamma}{A}}+\frac{\frac{P^2}{A^2}}{8 \sqrt{\varpi^3}} \arcsin \frac{2 \varpi \Gamma-\frac{P}{A}}{\frac{P}{A}}\right)\right]_{\Gamma=\Gamma_{1}}^{\Gamma=\Gamma_{2}} .
    \end{align}
    \label{v1_Gamma}
    \end{subequations}
\end{small}
    \hrulefill
\end{figure*}


\section{Simulations}
In this section, we provide simulation results to illustrate the fact that the performance gain of ISAC is determined by the subspace correlation. Without loss of generality, we consider an ISAC BS that is equipped with $N_t = 10$ and $N_r = 12$ antennas with its transmitter and receiver and the antenna spacing is half wavelength. The ISAC frame length is set as $T = 30$. The maximum transmit power is set as $P = 20$ dBm and the noise power is set as $\sigma_c^2 = \sigma_r^2
= 0$ dBm. The target is located at $\theta = 15^{\circ}$. The communication channel $\mathbf{h}_{\mathrm{c}}$ is set as Rayleigh fading and each entry follows $\mathcal{CN}(0,1)$ distribution. The integral interval in \eqref{integration} is set as $\Gamma_1 = 0.4~\Gamma_{\max}$ and $\Gamma_2 = 0.95~ \Gamma_{\max}$, where $\Gamma_{\max}$ denotes the maximum achievable SNR computed by $P\|\mathbf{h}_{\mathrm{c}}\|^2/ \sigma_c^2$.

In Fig. \ref{Fig1a}, we show the closed-form and numerical solutions for AO criterion and Det-Max criterion. The numerical solution of AO and Det-Max criteria are computed through numerically solving the convex optimization problem \eqref{P_AngleOnly} and \eqref{P1_Det} by CVX, respectively, while the closed-form solution of the two criteria is presented in \eqref{opt_Rx_AngleOnly}. As illustrated in Fig. \ref{Fig1a}, the closed-from solutions match well with their numerical counterparts, both for AO and Det-Max criteria. Moreover, the increase of the required communication rate threshold at the CU leads to rising root-CRB of the target angle and reflecting coefficient.

Fig. \ref{Fig1b} indicates that the optimal beampatterns associated with two criteria are nearly the same. Moreover, we choose two benchmarks, namely CRB-minimization and communication-optimum criteria. The communication-optimum beampattern is attained by only maximizing the achievable rate while ignoring the sensing performance. The CRB-minimization beampattern is attained by solving problem (23) in \cite{lijian2007range}. It is shown that the peak of the beampattern around the target location in ISAC is lower than that in the CRB-minimization criterion but higher than the communication-optimum criterion. This result strikes a flexible tradeoff between both CRB-minimization and communication-optimum beampatterns to balance S$\&$C.

\begin{figure}[!t]
    \centering
    \epsfxsize=1\linewidth
    \includegraphics[width=8cm]{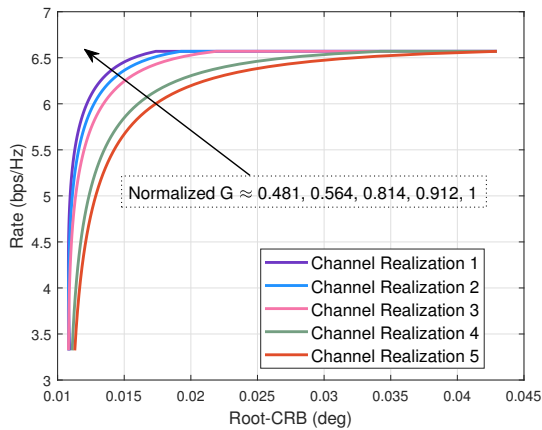}
    \caption{ISAC performance region characterized by the normalized ``correlation coefficient'' $\mathsf{G}$.}
    \label{fig2}
\end{figure} 

In Fig. \ref{fig2}, we characterize the ISAC performance region by the pre-defined ``correlation coefficient'' $\mathsf{G}$ over several realizations of communication channels. We normalize each randomly generated channel vector with $\mathbf{h}_{\mathrm{c}}^i = \mathbf{h}_{\mathrm{c}}^i / \|\mathbf{h}_{\mathrm{c}}^i\|, i = 1,2,\ldots,5$, which consequently ensures a fair comparison that the integration gain originates from the correlation between S$\&$C subspaces instead of the communication channels. Ranging from the Channel Realization 1-5, the results show that the ISAC system with a higher ``correlation coefficient'' acquires better performance in both the CRB and achievable rate, which suggests that the integration gain of the ISAC system may be measured by the ``correlation coefficient'' defined in \eqref{Def_G}. 

\section{Conclusion}
In this paper, we revealed that the integration gain of ISAC systems is determined by the subspace correlation between S$\&$C channels. We first showed that the ISAC subspace is spanned by the transmitted steering vectors of the sensing channels and the right singular matrix of the communication channels. Then, we derived the closed-form solutions of the optimal transmitted waveform both in angle-only and determinant-maximization criteria. The insight investigated from the closed-form solution is that if the defined subspace ``correlation coefficient'' of S$\&$C is larger, the ISAC system will attain higher performance gain. Finally, the simulation results illustrated the effectiveness of the proposed approaches. 

\bibliographystyle{IEEEtran}
\bibliography{ref}

\end{document}